\documentclass[a4paper]{jpconf}
\usepackage{graphicx}
\usepackage{amssymb}

\def\Journal#1#2#3#4{{#4}, {#1}, {#2}, #3} 

\begin{document}
\title{AMS-02 in Space: Physics Results}

\author{Nicola Tomassetti, on behalf of the AMS collaboration}
\address{LPSC, Universit\'{e} Grenoble -- Alpes, CNRS/IN2P3 -- Grenoble, France}

\ead{nicola.tomassetti@lpsc.in2p3.fr}

\begin{abstract}
The Alpha Magnetic Spectrometer (AMS-02) is a particle physics experiment 
designed to study origin and nature of Galactic Cosmic Rays (CRs) up to TeV energies from space. 
With its high sensitivity, long exposure and excellent identification capabilities, 
AMS is conducting a unique mission of fundamental physics research in space. 
To date, more than 60 billion CR events have been collected by AMS. 
The new results on CR leptons and the analysis and light-nuclei are presented and discussed.
The new leptonic data indicate the existence of new sources of high-energy CR leptons,
that may arise either by dark-matter particles annihilation or by nearby astrophysical sources of $e^{\pm}$ pairs. 
Future data at higher energies and forthcoming measurements on the antiproton spectrum and the boron-to-carbon 
ratio will be crucial in providing the discrimination among the different scenario.
\end{abstract}

\section{Introduction}     
\label{Sec::Introduction}  

The Alpha Magnetic Spectrometer (AMS) is an international project devoted to
search for new physics phenomena, such as primordial antimatter or dark matter, 
by means of direct detection of high-energy cosmic-ray (CR) particles in space.
The final version of the detector, AMS-02, was successfully installed and activated 
on the International Space Station on May 19$^{\rm th}$ 2011. 
AMS-02 is capable of measuring CR leptons and nuclei, from hydrogen up to iron, from hundreds MeV up to $\sim$\,1\,TeV of energy,
with unprecedent precision and sensitivity.
The AMS-02 data are expected to significantly improve our understanding of the CR acceleration and propagation processes in the Galaxy.
The status of AMS measurements of CR leptons and light-nuclei is reviewed in these proceedings.
%
The layout of the AMS-02 instrument and its complete description can be found in earlier papers \cite{Bib::AMSPositronFraction2013}.
The detector is composed of a Silicon Tracker, a Transition Radiation detector (TRD), a Time of Flight system (TOF), 
a permanent magnet, an array of AntiCoincidence Counters (ACC), a Ring imaging Cherenkov detector (RICH), 
and an Electromagnetic Calorimeter (ECAL).
Redundant measurements ensure a complete particle identification and allow for a careful study of interactions inside the detector
\cite{Bib::TomassettiOliva2013, Bib::Saouter2013}.
The TRD \cite{Bib::TRD} allows for lepton/hadron separation. It
is made of 5248 proportional chambers, filled with xenon and carbon-dioxide,
arranged in 20 layers of fiber--fleece radiators.
Four TOF planes of scintillators \cite{Bib::TOF}, located above and below the magnet case, 
allow for velocity measurements, charge measurements, and trigger.
The Tracker is build of nine layers of silicon micro-strip detectors \cite{Bib::Tracker}, from L1 to L9, 
distributed over the instrument. Layers from L2 to L8 are assembled with the permanent magnet, that has a magnetic field
strength of 0.15\,T. L1 is located on top of the TRD, and L9 is located between between RICH and ECAL. 
The Tracker measures the rigidity $R$ up to $R\approx$2 TV ($R\equiv p/Z$, momentum to charge ratio) 
for charge one particles, and charge measurement up to $Z=28$.
The RICH \cite{Bib::RICH} is made of a radiator layer, a conical mirror, and a photomultiplier plane 
which detects Cherenkov light. 
Two radiators, with  NaF and aerogel (refractive indices $n$=1.34 and $n$=1.04), are used to provide
high-precision measurement of the particle velocities and charge.
The ECAL \cite{Bib::ECAL} is made of lead and scintillating fibers arranged in 16 layers,
providing 17 radiation lengths of detecting medium. This allows for a precise reconstruction 
of the energy up to several TeV. A boosted decision tree classifier (BDT) is used to identify leptons
based on the topology of their showers.
Onboard the ISS, AMS-02 is orbiting the Earth at an altitude of about 400 km and with inclination of 51.6$^{\circ}$.
The average trigger rate is about 600 Hz,  the average event size is 2 kByte. The minimal down-link bandwidth is 9 Mbit/s.
AMS-02 collects about 1.5 billion CRs during each month of operation. 
The detector is controlled from the AMS-02 Payload Operations Control Center (POCC) at CERN, Geneva. 
Intensive time-dependent calibrations have been performed to all sub-detectors. 
No significant degradation of the sub-systems has been observed during 3 years of operation in the ISS.

\section{Results on cosmic-ray leptons} 
\label{Sec::Leptons}                    

The new results on CR leptons are based on the data collected during the initial 
3 years of operations on the ISS (May 2011 -- May 2014), which is about 16\% of the total expected data sample.
In the measurement of leptons, the TOF is used to select relativistic particles traversing AMS-02 in the downward 
direction. The signals in TRD and ECAL are used to discriminate the leptonic component from the hadronic background. 
A track reconstructed from L2 to L8 and matching the TRD and ECAL signals is used to select clean $Z = 1$ particles. 
\begin{figure*}[htbp]
\centering
\includegraphics[width=0.43\textwidth]{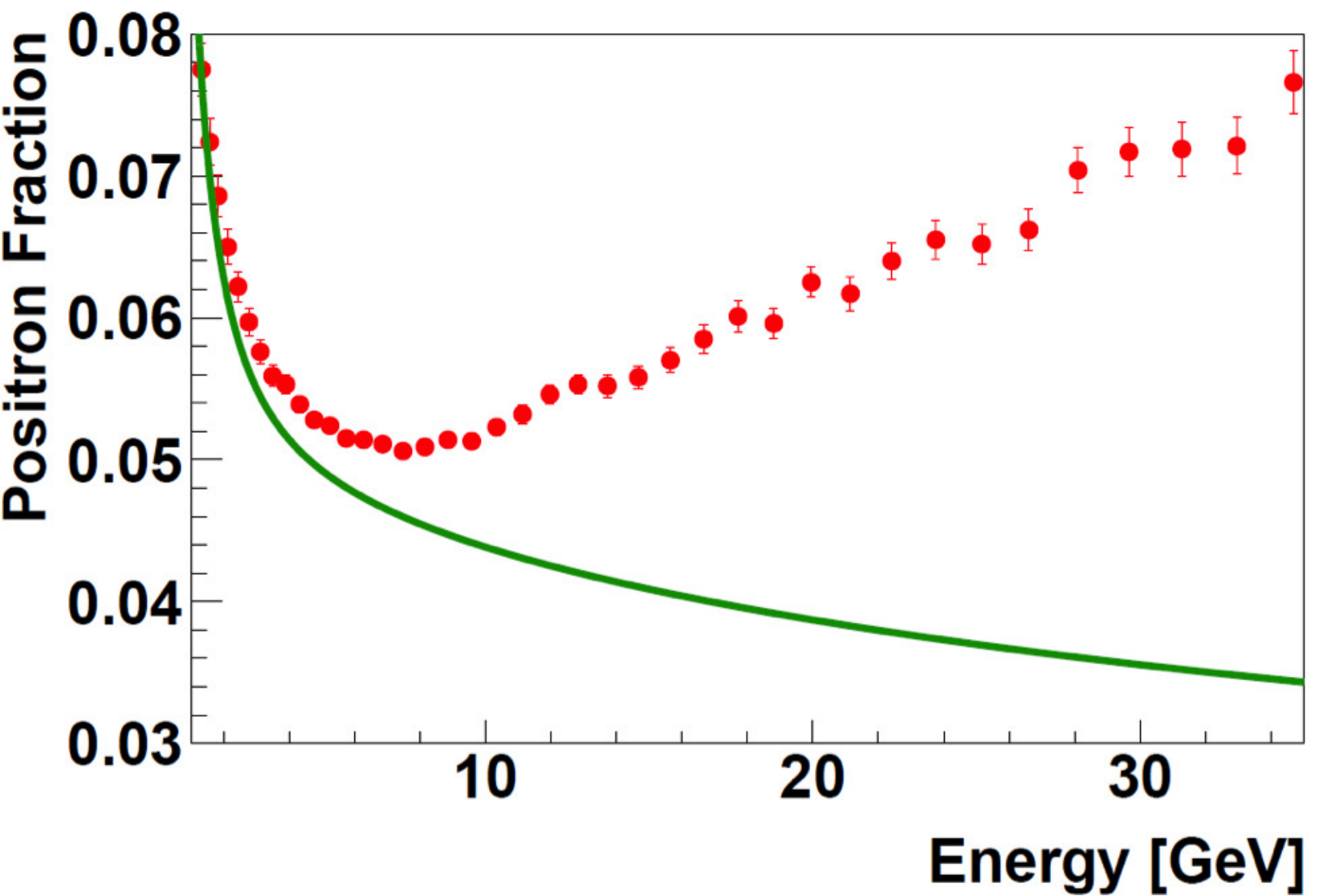}
\includegraphics[width=0.48\textwidth]{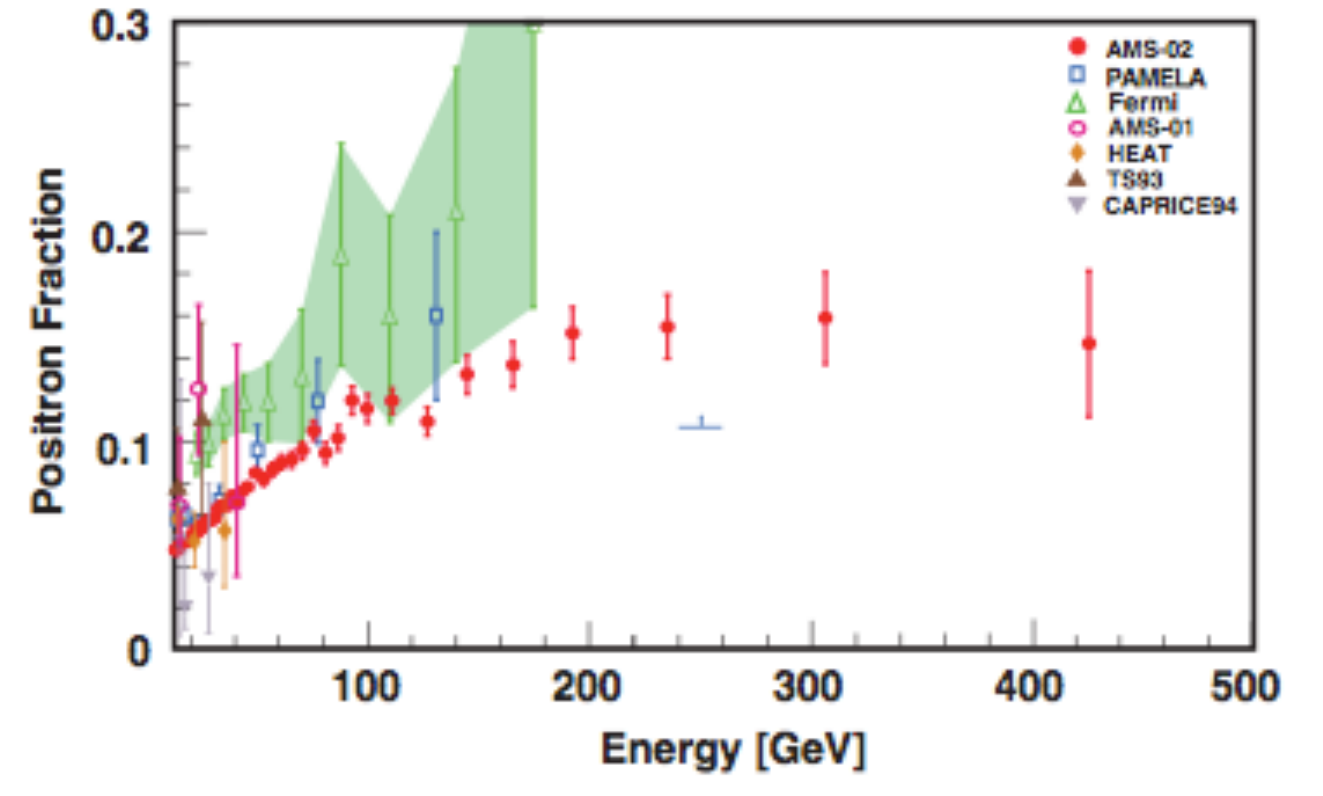}
\caption{AMS-02 positron fraction data from 1 to 35 GeV (left) and from 10 to 500 GeV (right) compared with previous measurements.}
\label{Fig::ccPositronFraction2014}
\end{figure*}
For the lepton/hadron separation, three methods are used:
a TRD-likelihood estimator, an ECAL-BDT estimator, and the $E/R$ ratio by combining ECAL and Tracker.
This redundancy allowed to study the performance of each discrimination method with the data. 
Various analyses with different combinations of cuts or template fits were also tested as a cross-check.
An important source of background comes from \emph{charge-confusion} of electron events, arising from
the finite rigidity resolution of the Tracker, or from the emission of secondary particles.
The charge-confusion contribution is studied from data and Monte Carlo simulations \cite{Bib::AMSPositronFraction2013}.
A high-energy measurement of the positron fraction (from 1 to 500 GeV) has been recently 
reported \cite{Bib::AMSPositronFraction2014}. The results are shown in Fig.\ref{Fig::ccPositronFraction2014}.
The left panel shows the fraction up to 35 GeV of energy compared with the standard conventional calculations
where CR electrons are emitted from supernova remnants (SNRs) and secondary $e^{\pm}$ arise from collisions of 
CR nuclei with the interstellar matter (ISM). 
This model predicts that the positron fraction decreases with energy. 
Above 10\,GeV, the AMS data show a persistent rise up to 200 GeV, in clear contrast with the conventional picture, 
followed by an intriguing flattening at higher energies (right panel).

\section{Results on cosmic-ray nuclei} 
\label{Sec::Nuclei}                    

Light nuclei with $Z>3$ constitute $\sim$\,1\% of the total CR flux.
\begin{figure*}[htbp]
\centering
\includegraphics[width=0.36\textwidth]{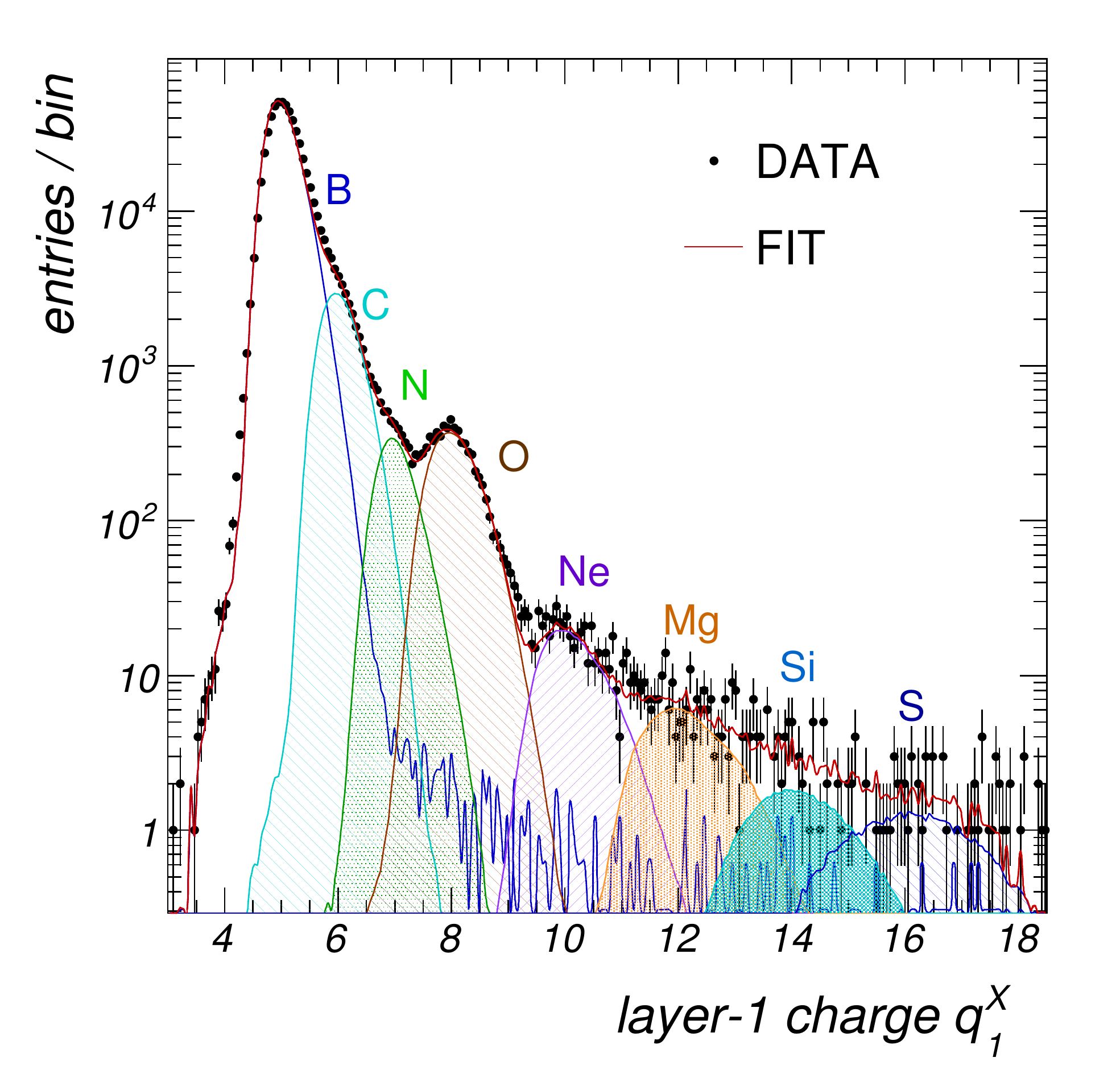}
\quad
\quad
\includegraphics[width=0.49\textwidth]{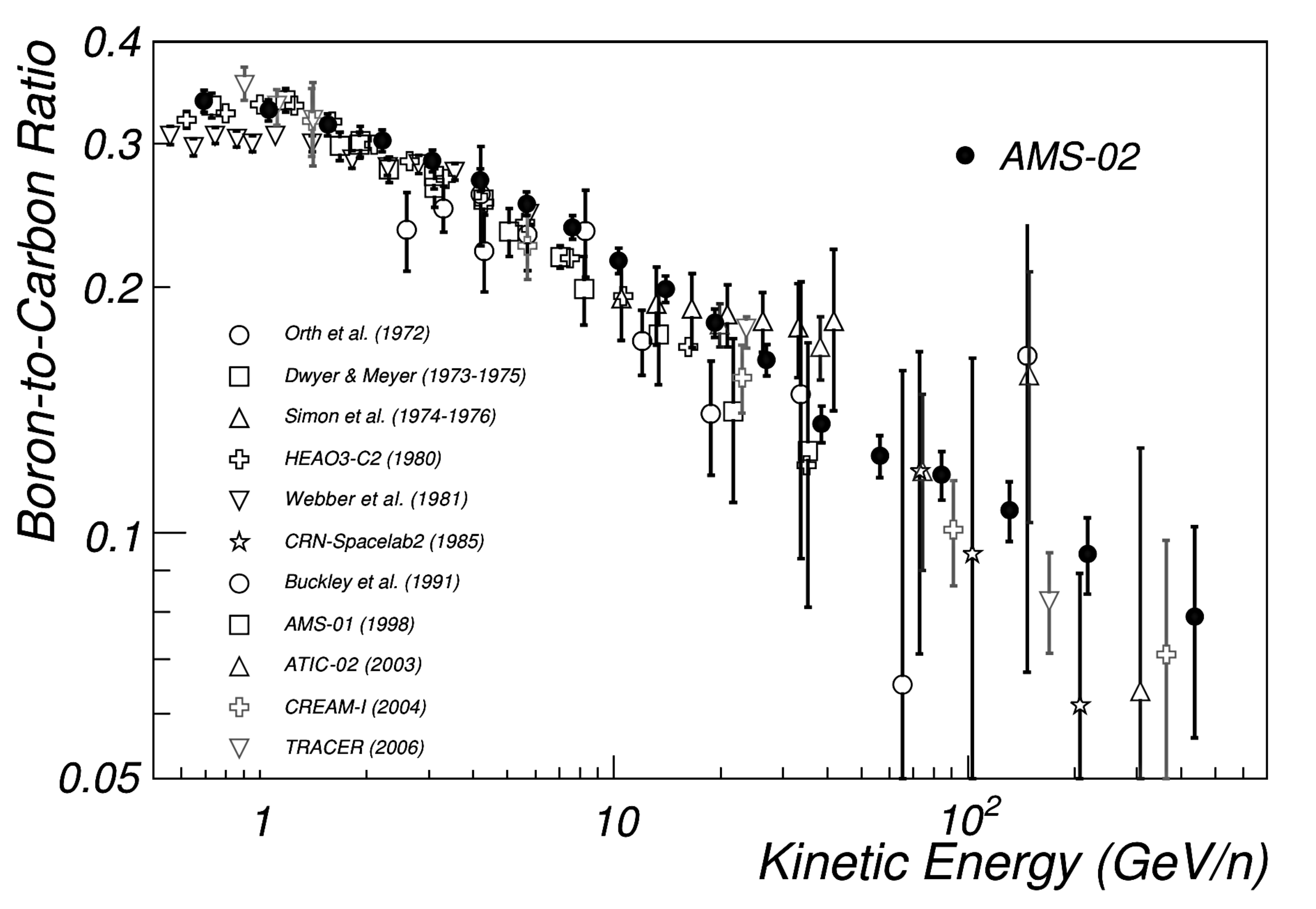}
\caption{
Left: Layer-1 Charge distribution for a loose Boron selection on TOF and Tracker (L2-L8). 
Fragmentation from heavier nuclei interacting in material below L1, such as C$\rightarrow$B or O$\rightarrow$B, 
can be estimated by a fit.
Right: the measured B/C ratio from 0.5 to 700 GeV/n \cite{Bib::Oliva2013}. 
}
\label{Fig::ccBCRatioICRC2013}
\end{figure*}
Rare elements such as $^{2}$H, $^{3}$He and Li-Be-B are believed to be of secondary origin, i.e. produced by collisions primary 
CRs (p, $^{4}$He, C-N-O, or Fe) with the gas nuclei of the ISM. 
Secondary to primary ratios such as $^{3}$He/$^{4}$He, B/C, or F/Ne are used to discriminate among astrophysical models of CR 
propagation in the Galaxy \cite{Bib::Strong2007}.
Furthermore, the B/C ratio at $\gtrsim$\,100\,GeV/n represents a crucial tool to understand 
open problems in CR physics such as the observed structures in primary CR spectra \cite{Bib::Tomassetti2012a}
or possible SNR origin for the positron excess \cite{Bib::MertschSarkar2014,Bib::TomassettiDonato2012,Bib::TomassettiDonato2015}.
The measurement of the B/C ratio is a prime physics goal of AMS-02. 
The analysis of the first 24 months of data has been presented in \cite{Bib::Oliva2013}.
With AMS-02, high purity samples of B and C events have been obtained using the TOF and Tracker
charge estimators, with a charge selection efficiency of 98\% and a charge mis-identification probability less than 10$^{-4}$.
Monte-Carlo simulated events are produced using \textsf{GEANT4-4.9.4} package and \textsf{DPMJET-II.5} \cite{Bib::Agostinelli2003,Bib::Ranft1995},
including a description of detector geometry, electromagnetic and hadronic physics processes, 
as well as the full digitization chain in order to reproduce the signals of all sub-detector.
The several charge detectors above and below the inner core of the spectrometer have been used to improve the 
overall charge separation capability as well as to select clean data samples of non-interacting CR particles. 
In particular, they allows for a data-driven characterization of nuclear interactions occurring at different 
levels in the spectrometers. Data and Monte-Carlo show an agreement at $\mathcal{O}(\%)$ level.
Figure\,\ref{Fig::ccBCRatioICRC2013} (left) shows the L1 charge distribution obtained after selecting pure B samples in 
Inner-Tracker (L2-L8) and TOF. The colored curves are the single-layer charge responses for known elements. 
The data (black markers) are described by a combination of these templates using Boron (94\%), Carbon (4\%), 
and heavier elements ($\sim$\,1\%). These numbers represent the probability for top-of-instrument processes
of charge-changing fragmentation, obtained with flight data. In this figure, no cuts are applied to the TRD signal. 
Simple cuts on the TRD or L1 charge allows to highly-efficiently suppress these events. 
The measured B/C ratio is shown in Fig.\,\ref{Fig::ccBCRatioICRC2013} (right) as a function of the kinetic energy obtained from the Tracker rigidity. 
Effect of bin-to-bin energy migration due to the finite resolution of the instrument has been accounted using an unfolding procedure.
The ratio decreases smoothly with energy. The B/C behavior at TeV energies will be become more transpaent with more data.

\section{Conclusions and discussion} 
\label{Sec::Discussion}              

New AMS results on the positron fraction show persistent rise up at $\sim$\,10 -- 200 GeV of energy, 
in clear contrast with conventional model predictions, followed by an intriguing flattening at higher energies.
From the measurement of the single spectra of $e^{+}$ and $e^{-}$ \cite{Bib::AMSElectronAndPositron2014},
the data show that both components $e^{-}$ and $e^{+}$ harden smoothly at $\gtrsim$\,30\,GeV.
These data are also consistent with the measured ``all-electron'' spectrum, $(e^{-} + e^{+})$, 
which has been independently determined up to $E=$\,1\,TeV \cite{Bib::AMSAllElectron2014}. 
All these data indicate the existence of a new high-energy source of $e^{+}$ and $e^{-}$, which
can be interpreted either in terms of dark matter (DM) particles annihilation
or in terms of nearby astrophysical sources of $e^{\pm}$.
The DM interpretation require annihilating particles with mass of the order of $\sim$\,1\,TeV, and may predict possible 
signals in the $\bar{p}/p$ ratio, depending on the specific DM--DM annihilation channels \cite{Bib::Boudad2014, Bib::Geng2014}. 
The existing antiproton data show no clear excess within the uncertainties in the data and in the model predictions.
In the second class, a recent work demonstrated that the AMS-02 data are described
well in terms of $e^{\pm}$ pairs emission from pulsar wind nebulae (PWNe) \cite{Bib::DiMauro2014}.
The PWN scenario gives no signatures in the antiproton channel.
It was also proposed that high-energy $e^{\pm}$ can be produced inside old-SNRs via interactions 
of CR protons with the background medium \cite{Bib::Blasi2009, Bib::MertschSarkar2014}. 
This mechanism predicts remarkable signatures in both the $\bar{p}/p$ ratio \cite{Bib::BlasiSerpico2009} and the $B/C$ ratio 
\cite{Bib::MertschSarkar2014}, that may be detectable by AMS-02 \cite{Bib::TomassettiDonato2012}. 
The data show no prominent signatures in the $B/C$ ratio \cite{Bib::Oliva2013,Bib::TomassettiDonato2015}.
The present models are affected by large astrophysical uncertainties 
that may be dramatically reduced with new data on CR nuclei \cite{Bib::Tomassetti2012b,Bib::Maurin2010}.

\section{Acknowledgement}    
\label{Sec::Acknowledgement} 

This work is supported by acknowledged persons and institutions 
in \cite{Bib::AMSPositronFraction2013} and by the LabEx grant \textsf{ENIGMASS}.
\\


\begin{thebibliography}{00}
\bibitem{Bib::AMSPositronFraction2013} Aguilar, M., et alii, \Journal{PRL}{110}{141102}{2013}
\bibitem{Bib::TomassettiOliva2013} Tomassetti, N., \& Oliva, A., 2013, 33$^{\rm th}$ ICRC, 896, Rio de Janeiro [arXiv:1510.09215]
\bibitem{Bib::Saouter2013} Saouter, P., et alii, 2013, 33$^{\rm th}$ ICRC, 789, Rio de Janeiro
\bibitem{Bib::TRD} Kirn, T., et alii, \Journal{NIM}{A 706}{43}{2013}
\bibitem{Bib::TOF} Basili, A., et alii, \Journal{NIM}{A 707}{99}{2013}
\bibitem{Bib::Tracker} Alpat, B., et alii, \Journal{NIM}{A 613}{207}{2010}
\bibitem{Bib::RICH} Aguilar, M., et alii, \Journal{NIM}{A 614}{237}{2010} 
\bibitem{Bib::ECAL} Adloff, C., et alii, \Journal{NIM}{A 714}{147}{2013} 
\bibitem{Bib::AMSPositronFraction2014} Accardo, L., et alii, \Journal{PRL}{113}{121101}{2014}
\bibitem{Bib::Oliva2013} Oliva, A., 2013, 33$^{\rm th}$ ICRC, 1266, Rio de Janeiro 
\bibitem{Bib::Strong2007} Strong, A. W., et alii, \Journal{Ann.Rev.Nucl.\&Part.Sci.}{57}{285--327}{2007}
\bibitem{Bib::Tomassetti2012a} Tomassetti, N., \Journal{ApJ}{752}{L13}{2012} [arXiv:1204.4492]
\bibitem{Bib::MertschSarkar2014} Mertsch, P., \& Sarkar, S., \Journal{PRD}{90}{061301}{2014}
\bibitem{Bib::TomassettiDonato2012} Tomassetti, N., \& Donato, F., \Journal{A{\&}A}{544}{A16}{2012} [arXiv:1203.6094] 
\bibitem{Bib::TomassettiDonato2015} Tomassetti, N., \& Donato, F., \Journal{ApJ}{803}{L15}{2015} [arXiv:1502.06150]
\bibitem{Bib::Agostinelli2003} Agostinelli, S., et alii, \Journal{NIM A}{506 3}{250--303}{2003}
\bibitem{Bib::Ranft1995} Ranft, J., \Journal{PRD}{51}{64}{1995}
\bibitem{Bib::AMSElectronAndPositron2014} Aguilar, M., et alii,\Journal{PRL}{113}{121102}{2014}
\bibitem{Bib::AMSAllElectron2014} Aguilar, M., et alii, \Journal{PRL}{113}{221102}{2014}
\bibitem{Bib::Boudad2014} Boudad, M., 2015, A\&A 575, A67 [arXiv:1410.3799]
\bibitem{Bib::Geng2014} Geng, C. Q., et alii, 2015, PRD 91 095006 [arXiv:1411.4450]
\bibitem{Bib::DiMauro2014} Di Mauro, et alii, \Journal{JCAP}{1404}{006}{2014}
\bibitem{Bib::Blasi2009} Blasi, P., \Journal{PRL}{103}{051104}{2009}
\bibitem{Bib::BlasiSerpico2009} Blasi, P., \& Serpico, P. D., \Journal{PRL}{103}{081013}{2009}
\bibitem{Bib::Tomassetti2012b} Tomassetti, N., \Journal{Astrophys.Space Sci.}{342}{131-136}{2012} [arXiv:1210.7355]
\bibitem{Bib::Maurin2010} Maurin, D., et alii, \Journal{A{\&}A}{516}{A67}{2010}

\end{thebibliography}
\end{document}